\documentclass[pre,aps,12pt]{revtex4}
\usepackage{amssymb,graphicx}

\newcommand{\D}{\mathbb{D}}

\newcommand{\Prob}{ \mbox{\rm Prob} }

\newcommand{\half}{{\frac{1}{2}}}

\newcommand{\La}{\left\langle}
\newcommand{\Ra}{\right\rangle}
\newcommand{\cut}[1]{}

\newcommand{\ix}{X}

\begin{document}                    

\title{ Learning Curves for Mutual Information Maximization
      }

 \author{R. Urbanczik}
 \affiliation{
            Institut f\"ur theoretische Physik,\\
            Universit\"at W\"urzburg, \\
            Am Hubland, \\
            D-97074 W\"urzburg, \\
            Germany \\
       }

\baselineskip 0.8\baselineskip

\begin{abstract}
 \setlength{\parindent}{0.0cm}
 An unsupervised learning procedure based on maximizing the mutual 
 information between the outputs of two networks receiving different
 but statistically dependent inputs is analyzed 
 (Becker and Hinton, Nature, 355, 92, 161). For a generic data model, 
 I show that in the large sample limit the structure in the data is
 recognized by mutual information maximization.
 For a more restricted model, where the networks are similar
 to perceptrons, I calculate the learning curves for zero-temperature
 Gibbs learning. These show that convergence can be rather slow, and
 a way of regularizing the procedure is considered.
\end{abstract}
\maketitle

\ 


\section{Introduction}

In unsupervised learning one often tries to find a  mapping $\sigma$
of a high dimensional signal $X$ into a simple output space
$\mathbb Y$ which preserves the interesting and important features
of the signal. The statement of the problem is rather vague and a
wealth of algorithm exist for the task which often define the 
meaning of "interesting and important" in terms of the algorithm itself
\cite{Dud01}. In search for a principled approach, it seems natural
to turn to information theory and to require that the mutual information
$I(X;\sigma(X))$ between the signal X and its encoding $\sigma(X)$ should be
large. Unfortunately, this is often a trivial problem. If one component
of $X$, say the first one, has a continuous distribution, 
the mutual information
between X and this component is infinite 
and so $I(X;\sigma(X))$ can be maximized
by simply choosing $\sigma$ to project $X$ onto its first component.

To arrive at a meaningful task one has thus considered maximizing 
$I(X;\sigma(X + \eta))$, where $\eta$ is isotropic Gaussian noise
\cite{Dec96}. Then
if $\sigma$ is constrained to be linear and X is Gaussian, 
the problem becomes equivalent to principal component analysis, 
but one can also consider nonlinear choices
for $\sigma$. The drawback of this approach is that if one reparameterizes
$X$, setting $\hat{X} = \psi(X)$, then maximizing 
$I(\hat X;\sigma(\hat X + \eta))$ will in general 
yield quite different results even
if $\psi$ is a simple linear and volume preserving mapping. So in this 
approach the meaning of interesting and important is implicitly defined by
the choice of a coordinate system for $X$.

It is much more natural to apply information theory when considering the
related scenario that one has access to two signals $X_1$ and $X_2$ which
are different but statistically dependent. For instance $X_1$ might be a visual
and $X_2$ the corresponding auditory signal. Then $I(X_1; X_2)$ is a 
reparameterization invariant measure of the statistical dependence of the two
signals and one can ask for a simple encoding of $X_1$ which preserves
the mutual information of the two signals. So in this scenario one
will look for a mapping $\sigma_1$ of $X_1$ into a simple output space
$\mathbb Y_1$ for which $I(\sigma_1(X_1); X_2)$ is large. 
This is the basic idea of the information bottleneck method
\cite{Tis99,Slo99}.

In the same setting, a more symmetric approach has been proposed by Becker
and Hinton \cite{Beck92,Beck94}. The idea is to look for simple encodings
$\sigma_1,\sigma_2$ of both signals which yield a large value of 
$I(\sigma_1(X_1);\sigma_2(X_2))$. An attractive feature of this approach
is that to compute the mutual information of the encodings one has
to estimate probabilities only in the simple output spaces $\mathbb Y_1$ and
$\mathbb Y_2$ and not in the high dimensional space of the signals 
themselves.

While the main thrust of this paper is to analyze Becker and Hinton's proposal 
using statistical physics, I shall first give some general characteristics of
what can be learned by maximizing $I$ for a large class of scenarios where
the approach seems suitable.
I then specialize to the case that the $\sigma_i$ are perceptron like 
architectures with discrete output values 
and setup a framework for analyzing learning from examples in the 
thermodynamic limit. Next, some learning curves obtained for specific
cases are discussed, and I conclude by addressing the limitation of the 
presented approach and some insights gained from it. 

\section{General Characteristics}

In general terms the mutual information of $X_1$ and $X_2$ is the 
KL-divergence between the joint distribution of $X_1$ and $X_2$ and
the product distribution of their marginal distributions. 
If the variables have probability densities this 
definition reads:
\begin{equation}
I(X_1; X_2) = \int {\rm d} x_1\, {\rm d} x_2\, p(x_1,x_2) 
\log_2\frac{p(x_1,x_2)}{p(x_1)p(x_2)}\,.   \label{minf}
\end{equation}
$I(X_1;X_2)$ is nonnegative and vanishes only 
if $X_1$ and $X_2$ are independent.
So a positive value indicates statistical dependence, 
and the ideal scenario for Becker and Hinton's proposal is that this 
dependence is such that for 
suitable functions $\tau_1$ and $\tau_2$ we have $\tau_1(X_1) = \tau_2(X_2)$
for any possible joint occurrence of a pair $(X_1,X_2)$. For instance,
$\tau_1(X_1)$ might be the common cause of the two signals.  I shall
further assume that the knowledge of $\tau_1(X_1)$ (or $\tau_2(X_2)$) 
encapsulates the entire statistical dependency of the two signals, so that the
joint density of $(X_1,X_2)$ can be written as
\begin{equation}
p(x_1,x_2) = 
 \frac{\delta_{\tau_1(x_1),\tau_2(x_2)}}{z_{\tau_1(x_1)}}
  p(x_1) p(x_2)\,. \label{distr}
\end{equation}
For brevity I have assumed that the $\tau_i$ take on discrete values, so
$\delta$ refers to Kronecker's delta and the normalization is
\begin{equation}
z_k \,=\, \Prob[\tau_1(\ix_1) \!=\! k] \,=\, \Prob[\tau_2(\ix_2) \!=\! k]\,. 
\label{balance}
\end{equation}

If the joint distribution of the signals is given by (\ref{distr}), it makes
sense to ask whether the $\tau_i$ can be inferred by observing only
$(X_1,X_2)$. This naturally leads one to consider the mutual information 
because a simple calculation shows that 
$I(\ix_1; \ix_2) = I(\tau_1(\ix_1);\tau_2(\ix_2))$. In the appendix I show, 
using standard information theoretic relations,
that any two mappings $\sigma_i$  which also preserve
the mutual information, 
$I(\ix_1; \ix_2) = I(\sigma_1(\ix_1);\sigma_2(\ix_2))$, are related to the
$\tau_i$ in a simple  way. Namely,
\begin{equation}
\tau_i(x_i) = \phi_i(\sigma_i(x_i)) \label{perm}
\end{equation}
holds identically for suitable mappings $\phi_i$, and in this sense the 
$\tau_i$ provide a simplest description of the data. If the $\sigma_i$
have the same number of output values as the $\tau_i$, the $\phi_i$ can
only be permutations. Of course, as an unsupervised learning procedure
maximizing $I(\sigma_1(\ix_1);\sigma_2(\ix_2))$, does not fix specific values
for the output labels. 
Despite of this, I shall sometimes call the $\tau_i$ teachers and 
take such trivial permutational symmetries into account only tacitly.

Realistically,
one will not be able to choose the $\sigma_i$ based on the 
knowledge of the entire distribution of $(X_1,X_2)$, but only have
access to a training set $\D$ of finitely many example pairs 
$(X_1^\mu,X_2^\mu)$ sampled independently 
from $(X_1,X_2)$. For a given $\sigma = (\sigma_1,\sigma_2)$, a pair of 
students, 
one will then compute the empirical frequencies
\begin{equation}
p_{u_1,u_2}(\D,\sigma) = 
  \frac{1}{m} \sum_{\mu=1}^m \prod_{i=1}^2 \delta_{u_i,\sigma_i(\ix_i^\mu)},
\label{empfreq}
\end{equation}
where $m$ is the number of examples in $\D$. Then the discrete version 
of (\ref{minf}) allow us to determine the empirical
mutual information $I(\D,\sigma)$ of the student pair on the training set by
\begin{equation}
I(\D,\sigma) = \sum_{u_1,u_2 = 1}^K 
p_{u_1,u_2}(\D,\sigma) 
\log_2\frac{p_{u_1,u_2}(\D,\sigma)}
       {p_{u_1,.}(\D,\sigma)p_{.,u_2}(\D,\sigma)}\, ,\label{empmimf}
\end{equation}
here $K$ is the number of output classes and the explicit formula for
the first marginal in (\ref{empmimf}) is
$p_{u_1,.}(\D,\sigma) = \sum_{u_2=1}^K p_{u_1,u_2}(\D,\sigma)$.

When learning, one has to restrict  $\sigma_1$ and $\sigma_2$
to lie in a  predefined set of functions
and the obvious strategy is to 
choose a pair maximizing $I(\D,\sigma)$. Of course,  Eq. (\ref{perm}) 
will then only hold in the limit $m\rightarrow\infty$
of an infinite training set, and a key issue is to quantify the speed of 
this convergence.
This seems especially important
since the number of values taken on by the $\tau_i$ is in general not
known. 
So it is quite possible that $K$ is chosen too large. Then,
even in the infinite training set limit, there can be
many different function pairs where $\sigma_i$  takes on all of the $K$ values,
$I(\sigma_1(\ix_1);\sigma_2(\ix_2))$ is maximized, but 
$\phi_i(\sigma_i(\ix_i)) = \tau_i(\ix_i)$ can satisfied by mappings
$\phi_i$ which merge class labels.
Thus  one will not expect 
that the number of classes in the data is automatically inferred
by mutual information maximization and  will have to experiment with
different values of $K$, considerably increasing the risk of over-fitting.

\section{Statistical Physics}

Let us now assume that the $\tau_i$ are  
perceptrons which yield output values in $0,\ldots,K-1$, 
and each $\tau_i$ is characterized by an $N$ dimensional weight vector 
$B_i$ of unit length and scalar biases $\kappa_i^k$, $k = 1,\ldots,K-1$. 
On an $N$-dimensional input $\ix_i$ the output of $\tau_i$ then is
\begin{equation}
\tau_i(x_i) = \sum_{k=1}^{K-1} \Theta( B_i^T x_i - \kappa_i^k),
 \label{teach}
\end{equation}
where $\Theta$ is the $0,1$ step function. While  Eq. (\ref{teach}) is 
invariant w.r.t. permutations of the biases, for brevity, I shall always
assume that the bias terms are in ascending order 
($\kappa_i^k \leq \kappa_i^{k+1}$).
The marginal densities $p(x_1)$ and $p(x_2)$ which are used to define
the joint density of the data (\ref{distr}),
are assumed to have 
independent Gaussian input components with $0$ mean and unit variance.
Then, to satisfy condition (\ref{balance}), the biases of $\tau_1$ and 
$\tau_2$ must be equal, 
$\kappa_1^k = \kappa_2^k = \kappa^k$. 

We assume that
the general architecture of the teachers is known, and
focus on pairs of students $\sigma_i$ performing a classification
analogous to Eq. (\ref{teach}) but with weight vectors $J_i$ and 
biases $\lambda_i^k$. Note that while formally I assume that the number
of biases is the same for teachers and students, this does not
restrict generality. For instance, a scenario where the teachers have fewer
output classes than the students is obtained by choosing some of the 
$\kappa^k$ to be equal.

The performance of a student pair is then assessed using (\ref{empmimf})
to determine $I(\D,\sigma)$.
To investigate, in the thermodynamic limit,  the typical properties of
maximizing $I(\mathbb{D},\sigma)$,
one has to fix a prior measure on the parameters of
the students. For the weight vectors, we  assume that the $J_i$ are drawn
from the uniform density ${\rm d}J$ on the unit sphere. 
As there are only finitely many $\lambda_i^k$ the results for 
$N\rightarrow\infty$ do not depend on the prior density ${\rm d}\lambda$
on the biases as long as the density vanishes nowhere.
One could now consider the partition function
\begin{equation}
Z = \int\!\! {\rm d}J\!\!  \int\!\! {\rm d} \lambda \: 
e^{\beta N I(\mathbb{D},\sigma)}
\end{equation}
for the 
Gibbs weight  $e^{\beta N I(\mathbb{D},\sigma)}$ on the space of students. 
But a key technical difference to many other learning paradigms is, that
this Gibbs weight does not factorize over the examples. There are, however,
some
special cases, namely if there are just two output classes and no biases, 
where one can
replace $I(\mathbb{D},\sigma)$ by an equivalent cost function which is just a
sum over examples \cite{Urb99}. Then maximizing $I(\mathbb{D},\sigma)$
is closely related to a supervised learning problem for parity machines.

Here, I want to analyze more general scenarios and it is easier not to start 
with  $e^{\beta N I(\mathbb{D},\sigma)}$ but to introduce 
target values $t_{u_1,u_2}$  for the empirical frequencies 
$p_{u_1,u_2}(\mathbb{D},\sigma)$ which determine 
$I(\D,\sigma)$. We now
consider the partition function
\begin{equation}
Z = \int\!\! {\rm d}J\!\!  \int\!\! {\rm d} \lambda \: 
\prod_{u_1,u_2}
     \exp\left( -\frac{\beta N }{2} 
           \left(t_{u_1,u_2} - p_{u_1,u_2}(\mathbb{D},\sigma)
                        \right)^2 
          \right)\;. \label{mypart}
\end{equation}
Analyzing the divergence of $\ln Z$ for $\beta\rightarrow\infty$, then 
tells us if the target values are feasible, i.e. whether student networks 
$\sigma_i$ exist with $t_{u_1,u_2} =  p_{u_1,u_2}(\mathbb{D},\sigma)$.

In the thermodynamic limit one will expect to find two regimes: As long as
the number of training examples $m$ is small compared to $N$, it will be 
possible to find students which achieve the global maximum $\log_2 K$ 
of the mutual information. 
In terms of the target values this means that 
$t_{u_1,u_2} = K^{-1} \delta_{u_1,u_2}$ is feasible, and we need to study
the partition function  (\ref{mypart}) for this choice of  $t_{u_1,u_2}$.
Once the ration $\alpha = m/N$ becomes large enough, there will in general 
be no 
students $\sigma$ such that $I(\mathbb{D},\sigma) = \log_2 K$ and we need
to determine the achievable empirical frequencies by finding feasible
target values of $t_{u_1,u_2}$ using Eq. (\ref{mypart}). We can then search for
the feasible target values which yield the  maximal mutual information
$I(\alpha)$.

For both regimes the starting point is to factorize (\ref{mypart}) over the 
patterns, linearizing the exponent 
by  an integral transform with Gaussians $L_{u_1,u_2}$ of 
$0$ mean and unit variance:
\begin{equation}
e^{-\frac{\beta N }{2} 
           \left(t_{u_1,u_2} - p_{u_1,u_2}(\mathbb{D},\sigma)
                        \right)^2 
  } =
\La e^{{\mathrm i}\, L_{u_1,u_2} \sqrt{\beta N}
     (t_{u_1,u_2} - p_{u_1,u_2}(\mathbb{D},\sigma))
}\Ra_{L_{u_1,u_2}}.
\end{equation}
One now employs standard arguments to calculate the quenched average
in the thermodynamic limit and finds, within a replica symmetric 
parameterization,
\begin{eqnarray}
\lim_{N\rightarrow\infty} N^{-1} \La \ln  Z \Ra_{\mathbb{D}} &=& 
\max_{R,\lambda} \min_{q,L} 
\alpha G_0(L) +  \alpha G_1(R,\lambda,q,L) + G_2(R,q), \nonumber \\
 G_0(L) &=&
  \sum_{u_1,u_2} \frac{L_{u_1,u_2}^2}{2 \beta} + L_{u_1,u_2}t_{u_1,u_2} 
\nonumber \\
 G_2(R,q) &=& \frac{1}{2}\sum_i\frac{q_i - R_i^2}{1-q_i} + \ln(1 - q_i)\,.
\label{lnZ}
\end{eqnarray}

Here $R_i = J_i^T B_i$ is the typical overlap with the teacher of a student
picked from the Gibbs distribution (\ref{mypart}) and $q_i$ is the squared
length of the thermal average of $J_i$. Further 
\begin{equation}
G_1(R,\lambda,q,l) = \La  
f_{\{ R_iq_i^{-\half}\} }(y_1,y_2)  \ln 
\sum_{u_1,u_2} e^{- L_{u_1,u_2}} \prod_i  H_{u_i}(\lambda_i,q_i,y_i)
\Ra_{y_1,y_2}
\end{equation}
where the $y_i$ are independent Gaussians with $0$ mean and unit variance. 
Further
\begin{equation}
f_{\{ R_iq_i^{-\half}\} }(y_1,y_2) =
\sum_{k} \frac{1}{z_k}\prod_i H_k(\kappa, R_iq_i^{-\half},y_i)
\end{equation}
with
\begin{equation}
H_{u_i}(\lambda_i,q_i,y_i) = 
H\left( \frac{\lambda_i^{u_i} - q_i y_i}{\sqrt{1-q_i}}\right) -
H\left( \frac{\lambda_i^{u_i+1} - q_i y_i}{\sqrt{1-q_i}}\right)\,.
\label{kgard}
\end{equation}
Here $H(z)$ is Gardner's $H$-function and to define Eq. (\ref{kgard}) for
$u_i = 0$ and  $u_i = K-1$, we adopt the convention that 
$\lambda_i^{0} = -\infty$ and $\lambda_i^{K} = \infty$. The definition 
of  $H_k(\kappa, R_iq_i^{-\half},y_i)$ is entirely analogous, also using
$\kappa^0 = -\infty$ and $\kappa^K = \infty$.

Note that the physical interpretation of the auxiliary variables
$L_{u_1,u_2}$ is that a student pair $\sigma$ picked from the
Gibbs density will yield empirical frequencies 
$p_{u_1,u_2}(\mathbb{D},\sigma) = t_{u_1,u_2} +  L_{u_1,u_2}/\beta$.
Reasonably, one will only consider target values $t_{u_1,u_2}$ for these 
frequencies which sum to $1$, and then the stationary values of 
$L_{u_1,u_2}$ must sum to $0$. This can of course also
be obtained by direct manipulation of Eq. (\ref{lnZ}).

We are mainly interested in evaluating (\ref{lnZ}) for 
$\beta\rightarrow\infty$.  
The stationarity conditions for the order parameters yield that 
the scaling of a conjugate
$L_{u_1,u_2}$ in this limit will depend on whether $t_{u_1,u_2}$ is
positive or zero. Denoting by $S_t$ the support of $t$, i.e. the set of 
pairs $u=(u_1,u_2)$ for which $t_{u_1,u_2} > 0$, the stationarity conditions
yield that  $L_{u_1,u_2}$ diverges with $\beta$ as $\ln \beta$ if 
$u \not\in S_t$. But for positive $t_{u_1,u_2}$, if $t$ is feasible,
$L_{u_1,u_2}$ diverges as  $-\ln \beta$,  while for two pairs 
$u, \hat u \in S_t$, the difference between the
conjugates
\begin{equation}
L_{u_1,u_2} - L_{\hat u_1,\hat u_2} = l_{u_1,u_2} - l_{\hat u_1,\hat u_2}
\end{equation}
stays finite for large $\beta$. Thus one obtains for
the limit $\beta\rightarrow\infty$
\begin{eqnarray}
\lim_{N\rightarrow\infty} N^{-1} \La \ln  Z \Ra_{\mathbb{D}} &=& 
\max_{R,\lambda} \min_{q,l} 
\alpha \!\sum_{u \in S_t}   l_{u_1,u_2}t_{u_1,u_2}  +  
\alpha \hat G_1(R,\lambda,q,l) + G_2(R,q)             \nonumber \\
\hat G_1(R,\lambda,q,l)   &=& 
\La  
f_{\{ R_iq_i^{-\half}\} }(y_1,y_2)  \ln 
 \!\sum_{u \in S_t} e^{-l_{u_1,u_2} } \prod_i  H_{u_i}(\lambda_i,q_i,y_i) 
\Ra_{y_1,y_2}.
\label{lnZ1}\nonumber \\ 
\end{eqnarray}
When the mutual information is maximized by marginally feasible target values
realized by only a single pair of students,  we need to
consider the limit $q_i\rightarrow 1$ in (\ref{lnZ1}). As usual, the
the sum over $u$ in $\hat G_1$ is dominated by its largest term in this 
limit.
Setting
\begin{eqnarray}
H^*_{u_i}(\lambda_i,y_i) &=& 
2 \lim_{q_i\rightarrow 1} (1-q_i) \ln H_{u_i}(\lambda_i,q_i,y_i) \nonumber \\
u^g(y_1,y_2) &=& 
{\renewcommand{\arraystretch}{0.7}
\begin{array}[t]{c}\mbox{argmax}\\ \scriptstyle u \in S_t\end{array}}\!\!
\left\{  g_{u_1,u_2} + 
         H^*_{u_1}(\lambda_1,y_1) + H^*_{u_2}(\lambda_2,y_2)/\gamma
 \right\} 
\end{eqnarray}
where, for $q_i\rightarrow 1$,  $\gamma$ is the ratio $\frac{1-q_1}{1-q_2}$
and $g_{u_1,u_2} = l_{u_1,u_2}(1-q_1)$,  one  obtains: 
\begin{eqnarray}
t_{u_1,u_2} &=& \La  f_{\{ R_i\} }(y_1,y_2) \delta_{(u_1,u_2),u^g(y_1,y_2)}
\Ra_{y_1,y_2}  \nonumber \\
1-R_i^2&=& -\alpha \La  
f_{\{ R_i\} }(y_1,y_2)H^*_{u_i^g(y_1,y_2)}(\lambda_i,y_i)  
\Ra_{y_1,y_2} \,. \label{boundary}
\end{eqnarray}
The interpretation of the above equations is that the target values
$t_{u_1,u_2}$ are marginally feasible  for some value of $\alpha$
if one can find $R_i$, $\lambda_i$, $g_{u_1,u_2}$  and $\gamma$ such 
that (\ref{boundary}) holds for $u_i = 0,\ldots,K-1$ and $i=1,2$. 

Using the above results, the learning curves for maximizing $I(\D,\sigma)$
in the large $N$ limit can be calculated. In the regime where 
$I(\alpha) = \log_2 K$ we use (\ref{lnZ}) with the target values 
$t_{u_1,u_2} = K^{-1} \delta_{u_1,u_2}$. But above a critical number of
examples $I(\alpha)$ will be smaller than $\log_2 K$. The  using 
(\ref{boundary}) to find the feasible targets $t_{u_1,u_2}$ which maximize
the mutual information, amounts to solving a constrained optimization
problem.

\newcommand{\blambda}[1]{ \lambda^{(#1)} }
\newcommand{\bkappa}[1]{ \kappa^{(#1)} }

\section{Learning Curves}

\begin{figure}
\includegraphics[scale=0.8]{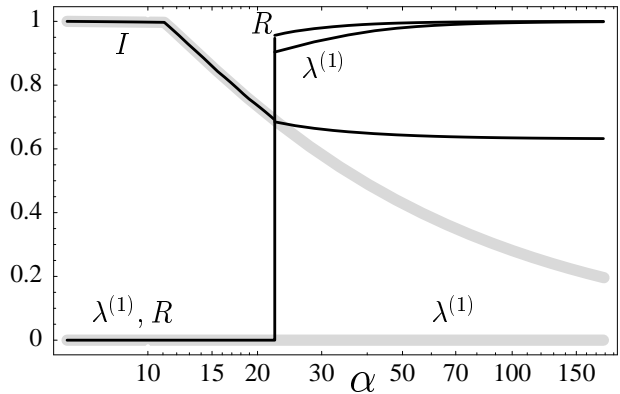}

 \caption{Learning curves for students with $K=2$ output classes. The  
  grey lines are for the random map problem, the thin black lines for a pair
  of teachers with two output classes and $\bkappa{1} = 1$.}
\end{figure}

Before considering example scenarios, some words on numerically 
solving Eqs. (\ref{lnZ1}) or (\ref{boundary}) are in order. This turns
out to be a non trivial task since averages of
functions have to computed which are quite non-smooth, once
the $q_i$ are close to $1$ in Eq. (\ref{lnZ1}), and become discontinuous for 
Eq. (\ref{boundary}).
To achieve reliable numerical results, I have found it 
necessary to explicitly divide the two dimensional domain of integration
into sub-regions where the integrand is both continuous and differentiable.
The number of sub-regions one has to consider increases quite rapidly with
$K$. 

Further, I have generally assumed site symmetry, 
$R_i = R,\, \lambda_i = \lambda,\, q_i = q$,
although I did numerically check the local stability of the solution thus
obtained for some points on the learning curves. 

The simplest case is that the students have $K=2$ output classes and
it is useful to first consider a 
degenerate scenario where the teachers have just
a single output. So $I(X_1,X_2)=0$ and the two signals are in fact
independent. 
This is analogous to the random map problem in supervised learning, since
nothing can be learned, and any pair of students will perform equally badly
on the whole distribution of inputs. But
for finite $\alpha$, up to $\alpha = 11.0$ , one can find student pairs 
achieving the maximal value $I(\D,\sigma)=1$, as shown in Fig. 1. 
Above this critical value the maximal 
empirical mutual information $I(\alpha)$ starts to decay to zero, 
the feasible target matrix $t$ 
becomes  non-diagonal but  the value of the bias $\blambda{1}$ is still zero. 
While above $\alpha=11.0$  student pairs with a diagonal $t$ do exist, and 
have a nonzero $\blambda{1}$, these pairs  do not maximize $I(\D,\sigma)$.

The random map problem is relevant for learning since the students always
have the option of ignoring the structure in the data. Formally, when $R=0$ 
a learning problem with $I(X_1,X_2)>0$ is equivalent to the $I(X_1,X_2)=0$
case.
This is illustrated (also Fig. 1) by a scenario where the teachers have
two output classes and $\bkappa{1} =1$.
This yields the moderate value $I(X_1,X_2) = 0.631$. 
But up to $\alpha=22.3$ the structure present in the data is 
not recognized at all and we observe the same behavior as for random 
examples. At  
$\alpha=22.3$ a first order phase transition occurs where $R$ and 
$\blambda{1}$ jump from zero to values which are already  close to $1$.

\begin{figure}
\includegraphics[scale=0.8]{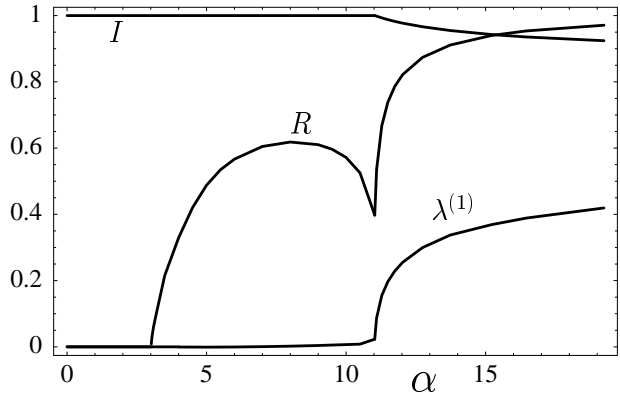}

 \caption{Learning curves obtained when the students and the pair of
          of teachers have two output classes but $\bkappa{1} = 0.5$.}
\end{figure}

When choosing $\bkappa{1} = 0.5$, still for $K=2$, a different behavior is 
observed 
since $I(X_1,X_2)$ is now quite close to $1$.
The phase where $I(\alpha) = 1$ is now  a bit longer, extending up to 
$\alpha = 11.1$. But already in this phase the order parameters show a non 
trivial behavior. The value of $R$ becomes positive above $\alpha=3.0$ but
is not monotonic in $\alpha$. So, while some structure is recognized in 
this phase due to entropic effects, the  recognition is rather unreliable. 
This is also highlighted  by the behavior of $\blambda{1}$. While it is
nonzero above $\alpha=3.0$, it initially even has very small negative values 
(not visible in Fig. 2). Above $\alpha=11.1$, when $I(\alpha) < 1$,
robust convergence of the order parameters to their asymptotic values sets 
in.

\begin{figure}
\includegraphics[scale=0.8]{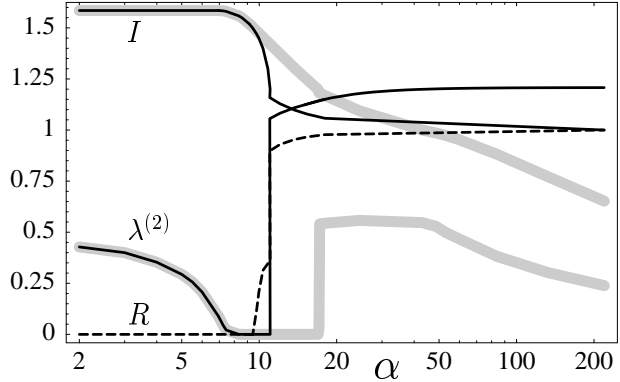}
 \caption{Learning curves for students with $K=3$ output classes. The  
  grey lines are for the random map problem, the black lines for a pair
  of teachers with three output classes and 
  $\bkappa{2} = - \bkappa{1} = 1.21$}
\end{figure}

Turning to $K=3$ (outputs $0$,$1$ or $2$), we again first consider the case
of random examples.
For all values
of $\alpha$ the bias term satisfies the symmetry $\blambda{2} = -\blambda{1}$.
The phase where
$I(\alpha)$ has the maximal possible value, which now equals $\log_2 3$, 
is shorter than for $K=2$, extending till $\alpha = 6.96$ as shown
in Fig. 3. 
Above $\alpha = 6.96$ the $t$-matrix is still diagonal initially. 

In this initial phase $\blambda{2}$ decreases with $\alpha$, this narrows
the gap between the output classes $0$ and $2$, making it easier to find a
student pair with $t_{02} = 0$. 
Remarkably, beyond $\alpha \approx 8$ one finds 
$\blambda{2} = \blambda{1} = 0$ but $t_{11} > 0$ as shown in Fig. 4.
This verges on the paradoxical since by definition a student 
with $\blambda{2} = \blambda{1}$ will never produce the output label $1$. 
However, we have taken the disorder average for $\blambda{1} < \blambda{2}$,
so the observed  result will naturally arise if the weight vectors of the 
optimal student pair satisfies $J_i^T X_i^\mu = 0$ on a subset of $\D$.
In addition, since we have take the thermodynamic limit first, 
$\blambda{2} = \blambda{1}$ may only hold in the large $N$ limit and not for 
finite $N$.

\begin{figure}
\includegraphics[scale=0.8]{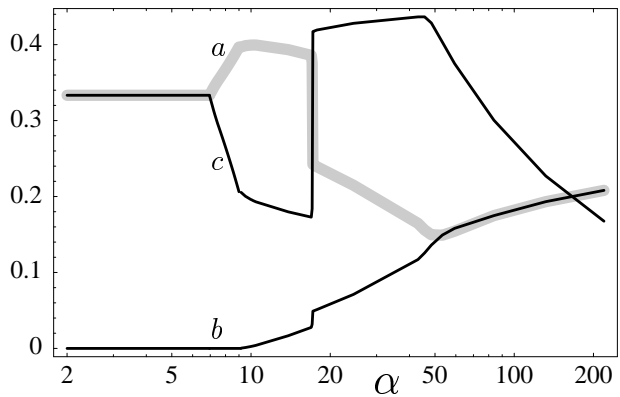}
 \caption{ Feasible $t$ values for $3$ output labels and random examples,
           $a = t_{00}, b = t_{02}, c = t_{11}$ as in Eq. (\ref{shape}).
}
\end{figure}

At $\alpha=9.2$ a continuous phase transition occurs with the $t$-matrix 
becoming non-diagonal (Fig. 4). It then has the form
\begin{equation}
t = \left(\begin{array}{ccc}
a&0&b\\0&c&0\\b&0&a\end{array}\right)\,. \label{shape}
\end{equation}
This is followed by first order phase transition at $\alpha = 17.2$ with 
$\blambda{2}$ jumping from $0$ to $0.55$. While the $t$-matrix keeps its shape 
(\ref{shape}), the values of  $c$ and $a$ change drastically.
The class of solution the network is now exploring, stays stable with 
increasing $\alpha$ and has a simple interpretation since 
the values of $a$ and $b$ converge.
This means that from the point of mutual 
information there is no difference between output $0$ and $2$. In effect 
the three output classes architecture is emulating perceptrons which have just
two output values but use
the non-monotonic output function $\Theta(\blambda{2} - |J_i^T\xi|)$.
While perhaps not quite as powerful as the reversed-wedge perceptron
 \cite{Bex95},
this architecture will have a very high storage capacity, and this leads to
a remarkably  slow convergence of $I(\alpha)$ to its asymptotic value of $0$.

The slow convergence for random examples suggests that it may be useful
to regularize
mutual information maximization and one way of doing this is considered in
Fig. 3. The  teachers have three output classes and 
biases $\bkappa{2} = - \bkappa{1} = 1.21$ yielding 
$I(X_1,X_2) =1$.
The students also have three output classes  but the training is regularized
by choosing students which maximize the mutual information under the
constraint that the $t$-matrix be diagonal, so the outputs of the two
students must be identical on the training set. 
The constraint becomes noticeable at $\alpha = 9.4$, where the achievable
$I(\alpha)$ is now lower than for the unconstrained case with $R=0$, i.e.
the random problem discussed above. Due to the constraint there is a 
continuous phase transition to positive $R$ at this point. 
Next, at $\alpha = 10.9$,  a first order phase transition to the 
asymptotic regime 
occurs, and the structure in the data is recognized well. At this point
the biases become nonzero and satisfy the symmetry 
$\blambda{2} = -\blambda{1}$.
Note that
up to $\alpha = 43$ the achievable $I(\alpha)$ is smaller than for
the unconstrained random map problem. So, regularizing the learning by
constraining the student outputs to be equal, is essential for
the good generalization observed for $\alpha$ values in the range
$[10.9,\ldots,43]$.

\section{Conclusion} 

We have seen that mutual information maximization provides a
principled approach to unsupervised learning. Interestingly,
from a biological perspective, it emphasizes the r\^ole of
multi-modal sensor fusion in perception. In contrast to many other
unsupervised learning schemes such as principal component analysis,
mutual information maximization can capture very complex
statistical dependencies in the data, if the architecture chosen for
the two networks is powerful enough.

For the generic data model given by Eq. (\ref{distr}), I have shown that
the structure in the data is recognized by mutual information maximization
if the training set is sufficiently large, i.e. the procedure is
consistent in a statistical sense. 
However, the detailed statistical physics calculations
yield  that many examples are needed to reach this asymptotic regime and
that the learning process is complicated by many phase transitions.
One reason for this is, that a seemingly simple architecture such as
a perceptron with three output classes can, from an information
theoretic point of view, be equivalent to a perceptron which has just two
output classes but uses a non-monotonic activation function.

Of course, when considering the number of examples needed for reliable
generalization, one has to keep in mind that
examples are often much cheaper in unsupervised than in 
supervised learning.  On the other hand, the detailed calculations have
been for cases, where the students are just perceptrons
and there are only few output classes. 
When increasing the number of output classes or when more
powerful networks are used, one will expect an even slower
convergence. So, in applications, it may be necessary to compromise the
generality of Becker and Hinton's approach by using suitable regularizations. 
We have considered one way of doing this, namely constraining the
two networks to give the same output on the examples in the
training set.

A major limitation of the above statistical physics analysis is
that I have only considered
the replica symmetric theory. It is, however, evident that in many of 
the above scenarios replica symmetry will be broken. A case in point is
the random map problem for two output classes where maximizing
the mutual information yields a critical value 
$\alpha=11.0$ up to which $I(\alpha) = 1$. This value is equal to the
storage capacity of the tree parity machine with two hidden units 
\cite{Bar90}, as one would expect, by the equivalence of the two problems in 
the unbiased case \cite{Urb99}. But one step of replica symmetry breaking,
considered in \cite{Bar90} for the tree parity machine, 
shows that the critical capacity is in fact some $25\%$ smaller.

To write down the one step symmetry breaking
equations for mutual information maximization, is a straightforward task. 
But given the numerical difficulties already
encountered in solving the replica symmetric equations, the numerics
of one step of replica symmetry breaking are daunting. While one will
expect that some of the quantitative findings described above change when
replica symmetry breaking is taken into account, one can reasonably assume
that more qualitative aspects such as the nature of the phase transitions
are described correctly by the present theory.

It is a pleasure to acknowledge many stimulating discussions with 
Georg Reents and Manfred Opper. This work was supported by the 
Deutsche Forschungsgemeinschaft.

\pagebreak
\appendix
\section*{Appendix}

Our goal is to show that if the joint density of $X_1$ and $X_2$ 
satisfies (\ref{distr}), then 
$I(\ix_1; \ix_2) = I(\sigma_1(\ix_1);\sigma_2(\ix_2))$ implies Eq. 
(\ref{perm}). We shall need  two facts from Information
Theory, see e.g. \cite{Cov91}. The first is the data processing inequality
(DPI), which states that for any mapping $\sigma$
\begin{equation}
I(\ix_1; \ix_2) \geq I(\ix_1; \sigma(\ix_2)),
\end{equation}
processing cannot increase information. The second is the chain rule which
allows one to decompose the mutual information of a random variable $X_1$
with a pair of random variables ($X_2,X_3)$ via:
\begin{equation}
I(\ix_1;\ix_2,\ix_3) = I(\ix_1;\ix_3) + I(\ix_1;\ix_2 \,|\, \ix_3 ),
\end{equation}
where the last term denotes the mutual information of the conditional
distribution of $(\ix_1,\ix_2)$ given a value of $\ix_3$, averaged over 
$\ix_3$.

Now, assuming Eq. (\ref{distr}), and
\begin{equation}
I(\ix_1; \ix_2) = I(\sigma_1(\ix_1);\sigma_2(\ix_2)) \label{labe}
\end{equation}
we have
\begin{eqnarray}
I(\ix_1; \ix_2) 
 &=& I(\ix_1; \tau_2(\ix_2),\sigma_2(\ix_2 )) \nonumber \\
 &=& I(\ix_1; \sigma_2(\ix_2)) +  I(\ix_1; \tau_2(\ix_2)\,|\,\sigma_2(\ix_2)) 
       \nonumber \\
 &=& I(\ix_1; \ix_2) +  I(\ix_1; \tau_2(\ix_2)\,|\,\sigma_2(\ix_2) )
\end{eqnarray}
Here the first equality is a consequence of the DPI and (\ref{labe}), 
the second is the chain rule, and the third is again DPI and  (\ref{labe}).

So $I(\ix_1; \tau_2(\ix_2)\,|\,\sigma_2(\ix_2) ) = 0$ and this means that
$\ix_1$ and $\tau_2(\ix_2)$ are conditionally independent 
given $\sigma_2(\ix_2)$. In other words:
\begin{equation}
p(\ix_1, \tau_2(\ix_2)\,|\,\sigma_2(\ix_2)) = 
p(\ix_1               \,|\,\sigma_2(\ix_2))\,\,
p(       \tau_2(\ix_2)\,|\,\sigma_2(\ix_2))
\end{equation}
or 
\begin{equation}
p(\ix_1, \tau_2(\ix_2),\sigma_2(\ix_2)) = 
p(\ix_1               ,\sigma_2(\ix_2))\,\,
p(       \tau_2(\ix_2)\,|\,\sigma_2(\ix_2))
\end{equation}
But from the definition of the joint density (\ref{distr}) we see
that $p(\ix_1, \tau_2(\ix_2),\sigma_2(\ix_2))$ can only be nonzero
if $\tau_1(\ix_1) = \tau_2(\ix_2)$ and in this case equals 
$p(\ix_1 ,\sigma_2(\ix_2))$. So $p(\tau_2(\ix_2)\,|\,\sigma_2(\ix_2))$
is either zero or one and this means that $\tau_2(\ix_2)$ is a function of 
$\sigma_2(\ix_2)$. By symmetry, this is also true of  $\tau_1(\ix_1)$ 
and $\sigma_1(\ix_1)$.

\bibliographystyle{unsrt}
\bibliography{../tex/neural}

\end{document}